\documentclass[twocolumn,superscriptaddress]{revtex4}
\usepackage{graphicx}
\usepackage{subfigure}
\usepackage{amsmath}
\usepackage{color}
\begin{document}
\draft
%twocolumn
%[\hsize\textwidth\columnwidth\hsize\csname@twocolumnfalse\endcsname
%\draft
\title{A numerical study of one-patch colloidal particles: from square-well to Janus.}
\author{Francesco Sciortino} 
\affiliation{ {Dipartimento di Fisica and CNR-ISC, Universit\`a di Roma {\em La Sapienza}, Piazzale A. Moro 2, 00185 Roma, Italy} }
\author{Achille Giacometti}
\affiliation{Dipartimento di Chimica Fisica, Universit\`a Ca' Foscari Venezia, Calle Larga S. Marta DD2137, I-30123 Venezia, Italy}
\author{Giorgio Pastore}
\affiliation{Dipartimento di Fisica dell' Universit\`a and CNR-IOM Democritos,
Strada Costiera 11, 34151 Trieste, Italy}
%\email{pastore@ts.infn.it}

\begin{abstract}
We perform numerical simulations of a simple model of one-patch colloidal particles to
investigate: (i) the behavior of the gas-liquid phase diagram on moving from a
spherical attractive potential to a Janus potential and (ii) the collective structure of a system
of Janus particles. We show that, for the case where one of the two hemispheres 
is attractive and one is repulsive, the system organizes  into
a dispersion of orientational ordered micelles and vesicles and, at low $T$, the
system can be approximated as a fluid of such clusters, interacting essentially via
excluded volume.  The stability of this cluster phase  generates a very peculiar
shape of the gas and liquid coexisting densities, with a gas coexistence density which
increases on cooling, approaching the liquid coexistence density at very low $T$. 
\end{abstract}
%61.20.Qg	Structure of associated liquids: electrolytes, molten salts, etc.
%81.16.Dn	Self-assembly
%82.70.Dd (colloids) , and 82.70.Gg (gels) 
%61.20.Ja Computer simulation of liquid structure

\maketitle

\section{Introduction}
The synthesis of colloidal particles with controlled anisotropy 
is central in today research.  One of the main ideas is the development of
a set of colloidal molecules\cite{mohovald,Manoh_03,Cho_05,kegel,simone,Glotz_Solomon_natmat,Paw10a} which can be useful to generate, on the nano and micron-scale, 
collective phenomena presently observed only at atomic or molecular scale as well as
additional  phenomena, induced by the possibility of controlling   the parameters of the 
effective interaction potential between colloids.  The self-assembly of a  colloidal diamond crystal, to be used in photonic applications\cite{fotonic}, is one  of these technological relevant goals.    The anisotropy can be induced not only by building colloidal molecules (i.e. colloids with peculiar non-spherical shapes), but also (with a  promising alternative) via the process of patterning the particle surface, generating in this way particles interacting in very different way according to their relative orientation\cite{mohovald, Janus,cacciuto,granich, janusgold,Stellacci}.  

This  vision of a material science, based on the design and self-assembly of materials with required properties has stimulated, beside the experimental work, a large amount of theoretical and numerical studies based on simple, primitive, anisotropic potentials\cite{Kern_03, bian, lungo, wilber-2006, doye, FoffiKern, whitelam}. Indeed, the possibilities of
modifying the surface properties are endless, including the number of patches, their width, their location, their chemical specificity,  and seem to pose no limits to the design of 
specific particle-particle interaction potentials.  In this large parameter space,  primitive
potentials --- modeling repulsion as a hard-core and attraction as a square-well interaction ---
can provide a useful reference system to deeply investigate the role played by the number of patches, their width, their spatial location and  the role of the  attraction range.
In addition, there is some consensus that these studies can shed light on the aggregation
properties of proteins and on the sensitivity of  the aggregates on the protein surface properties\cite{Sear_99,Lomakin,KumarJCP07,Li2009}.

Another useful aspect of the study of primitive models is the possibility to accurately investigate
theoretically and numerically  their phase diagram, casting the self-assembly process of these systems 
into a wider thermodynamic perspective. Important questions about the nature of the
self-assembly process and its reversibility, the relative stability of aggregates of different sizes, the competition between self-assembly and phase-separation can in principle be addressed 
with an accurate study of these models. Finally, in some cases,  numerical simulations can be compared with state of the art integral equation approaches for non-spherical potentials
providing a benchmark for the possibility of developing a fast and accurate 
prediction of the structural properties. Several efforts in this directions, capitalizing on 
studies of molecular associations\cite{Werth1,Werth2,Duda,Netzbeda2007B},  are taking place in these days\cite{millergiacometti,Giacometti09,kerntwo}.

In this article we report a study of the phase behavior of   a very simple primitive potential,
proposed in 2003 by Kern and Frenkel\cite{Kern_03} with the aim of  exploring
the scaling of the critical parameters (including the reduced second virial coefficient)
on the number and width of the patches,  and the possible implications for the phase behavior of globular proteins.
We limit ourselves to the case where the surface of the colloid particle is divided into two
parts, respectively repulsive and attractive.  On decreasing the surface area corresponding to the attractive part, the potential
interpolates between the well-known isotropic square well potential and  the symmetric case
of an evenly divided surface, commonly indicated as Janus potential. We find that
an unconventional phase diagram characterizes Janus particles, due to the onset of a micelle formation process which takes place in the gas phase, providing additional stability to this phase as compared to the liquid one.  We characterize the structural and connectivity properties of the system in a wide range of temperatures $T$ and number densities $\rho$, to clarify the origin of the aggregation process and the mechanisms behind the stability of the cluster phase.
Our study demonstrates how a small change in the attractive surface has profound consequences on the collective behavior of the system.

\section{Model and Simulation Techniques}
%% Achille: Ho cambiato leggermente le notazioni dei versori per essere consistenti con quanto appare nelle Equazioni (8) e (9) %%%

The Kern-Frenkel potential is a paradigmatic model for highly anisotropic interactions. In this model,  a hard-sphere of diameter $\sigma$
 is complemented by a set of unit vectors $\{\hat{\mathbf{n}_i}\}$,  locating the position of
the center of a  patch on the particle surface.   
Each patch can be reckoned as the intersection of the sphere with a cone of semi-amplitude $\theta$ and vertex at the center of the sphere. 
In the case studied in this model, each particle has only one patch. 
When unit vectors
%% Giorgio delete: the line joining the centers of two distinct particles $i$ and $j$ forms with both 
the  patch 
${\bf \hat{n}_i}$ of particle $i$ and ${\bf \hat{n}_j}$ of particle $j$
%% Giorgio insert:
form
%% /Giorgio
an angle smaller than  $\theta$,  and in addition the distance between the center of the two
particles is  between $\sigma$ and $\sigma+\Delta$, then an attractive interaction of 
intensity $u_0$ is present. More precisely, the two body potential is defined as:
%% Achille: ho tolto la parentesi grafa nella parte angolare perche' sembrava indicare che
%% dipendesse da tutti gli angoli invece che da uno solo e mi sembrava facesse confusione.

\begin{equation}
\label{eq1}
u(\mathbf{r}_{ij})=u^{sw}(r_{ij})f(\Omega_{ij})
\end{equation}
where $ u^{sw}(r_{ij})$ is an isotropic square well term of
depth $u_0$ and attractive range $\sigma+\Delta$ and
 $f(\Omega_{ij})$ is a function that depends on the orientation of
the two interacting particles $\Omega_{ij}$.  The angular function
$f(\Omega_{ij}) $ is defined as

\begin{equation}
\label{eq2}
f(\Omega_{ij}) = \left\{
  \begin{array}{ll}
1 &\mbox{if \, }\left\{
  \begin{array}{lll} \mbox{ $\hat{\bf r}_{ij}\cdot\hat{\bf n}_{i}>\cos
      \theta$ } & \begin{array}{l} \mbox{patch } \\\mbox{
        on particle $i$}\end{array}\\\mbox{and}&\\ 
    \mbox{ $\hat{\bf r}_{ji}\cdot\hat{\bf n}_{j}>\cos
      \theta$ } & \begin{array}{l}  \mbox{patch } \\\mbox{
        on particle $j$}\end{array}
            \end{array} \right.\\
0 & \mbox{else}
\end{array} \right.
\end{equation}

The diameter of the
particles and the depth of the square well have been chosen as units of
length and energy respectively, i.e.  $\sigma=1$ and $u_0=1$.

In practice, two particles interact attractively if, when they are within the
attractive distance $\sigma+\Delta$, two patches are properly facing each
other. When this is the case, the two particles are considered bonded. The fraction of
surface covered by the attractive patches $\chi$ is related to $\theta$ by the relation 
$\chi =  \frac{1-\cos \theta}{2}$. 
 
 Structural properties of the system have been evaluated by mean of standard MC simulations,
 for a system of $N=5000$ particles.  Extremely long simulations, of the order of $10^9$ MC sweeps,
 have been performed to reach a proper equilibrium state of the system. Here a MC sweep is defined as
 an attempted random translation and rotation for each particle.    To calculate the location of the gas-liquid critical point we perform grand canonical Monte Carlo (GCMC) simulations~\cite{frenkelsmith},
complemented with histogram reweighting techniques to match the
distribution of the order parameter $\rho - s e$ with the known
functional dependence expected at the Ising universality class
critical point~\cite{Wilding_96}. Here $e$ is the potential energy
density, $\rho$ the number density and $s$ is the mixing field
parameter. We did not performed a finite size study, since we are only
interested in the trends with $\chi$.  We have studied systems of 
different sizes, up to  $L=15$. For each studied $\chi$ --- using the
methods described in ~\cite{flavio} --- we calculated the critical
temperature $T_c$ and density $\rho_c$ for values of $\cos \theta$
between $-1$ and $0$ (at fixed $\Delta=0.5$).
Temperature is measured in reduced units, i.e. Boltzmann constant $k_B=1$. 

We also performed Gibbs ensemble simulations to evaluate  the coexistence curve.
The GEMC method was designed~\cite{GEMC} to study
coexistence in the region where the gas-liquid free-energy barrier is
sufficiently high to avoid crossing between the two phases.
Since nowadays this is a standard method in computational physics, we do
not discuss it here. We have studied a system of (total) 350 particles which
partition themselves into two boxes whose total volume is $2868\sigma^3$,
corresponding to an average density of $\rho=0.122$. At the lowest $T$ this
corresponds to roughly $320$ particles in the liquid box (of side $\approx 8\sigma$)
and about 30 particles in the gas box (of side $\approx 13\sigma$). Equilibration
at the lowest reported $T$ required about one month of computer time.

The model, in the case $\chi=0.5$, can be  related to the experimental system that
is currently under investigation\cite{granich}.  In these newly synthesized Janus particles,  
the repulsive interaction has an electrostatic origin and the  attractive part is hydrophobic
 At the present time, experiments by Granick's group\cite{granich}  have focused on the  analysis of the structure of the aggregates sedimented on the bottom surface due to gravity.  Micelle formation has been observed. While the interaction range in the Granick's experimental system  is about 0.1 of the particle size, nano-sized particles  synthesized with the same protocol would indeed give rise to potential ranges similar to the one we have selected.  It is also in principle possible to modify the range of the interaction  by tuning the physical properties of the solvent (for example the ionic strength or the solvent dielectric constant). Another interesting possible experimental realization of longer range interactions can be achieved  by mean of the recently measured Casimir critical forces [for a highlighting  review see for example Ref.~\cite{casimir1full}  or  Ref.~\cite{casimir2full}] dissolving the particles close to the critical point of the solvent. By tuning the distance from the critical point, the range of the interaction can be controlled and tuned close to the value we have explored.  Moreover, in the critical Casimir effect,  the coating of the two hemispheres will controls if the interaction is attractive or repulsive. Other possibilities of probing different ranges will be offered by studies of Janus-like proteins. Indeed, the hydrophobin proteins (extracted from fungi) are good candidates for nano-Janus particles, as discussed in Ref.~\cite{janus-softmatter}. Even in this protein case, the bulk phase diagram has never been explored. 

\begin{figure}[h]
\includegraphics[width=8cm, clip=true]{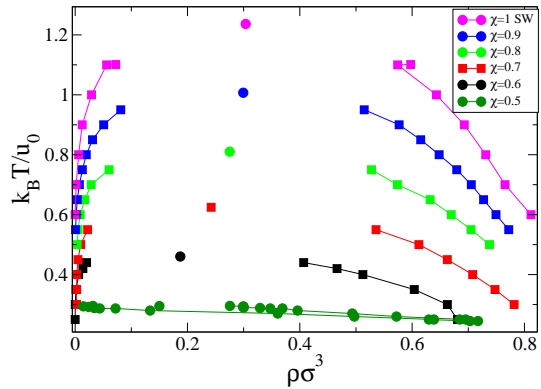}
\caption{Phase diagram of the one-patch Kern-Frenkel potential with attractive range $0.5 \sigma$ for different values of the coverage $\chi$, interpolating between the square-well potential ($\chi=1$) and the Janus potential ($\chi=0.5$).}
\label{fig:phase}
\end{figure}

\section{Results: From Square-well to Janus}

The Kern-Frenkel model offers the possibility to continuously change the
coverage interpolating from the isotropic square-well potential to the symmetric Janus-like one,
when the coverage moves from $\chi=1$ to $\chi=0.5$.  To investigate how the
phase diagram of Janus particles arises, we start by looking how the gas-liquid
coexistence is modified on progressively reducing $\chi$.   
Fig.~\ref{fig:phase} depicts the gas-liquid phase coexistence  for several $\chi$ values, extending the original
data by  Kern and Frenkel\cite{Kern_03}. 
While the
critical density  presents a significant decrease with decreasing $\chi$, the density of the liquid branch 
does not show any significant reduction, consistent with the possibility of forming 
six or more bonds with neighboring particles when $\chi>0.5$.   
Recent studies on the role of the valence (defined as the maximum possible number of bonded nearest neighbors)\cite{bian,23},  have suggested a progressive 
reduction of the critical temperature $T_c$ and critical density $\rho_c$ on decreasing the valence.
Indeed,  when the valence decreases below six, a significant reduction of the
density of the liquid branch has been reported\cite{lungo}. 
Estimates of the critical parameters have been calculated using grand-canonical simulations, i.e. simulations in which the chemical potential $\mu$,  $T$ and the volume $V$ are
kept constant. In GCMC simulations, the number of particles fluctuates. When $\mu$ and $T$ are close to their critical values, the number of particles fluctuates widely (since the compressibility, which is  a measure of the variance of the density fluctuations diverges). The distribution of
sampled densities, close to the critical point, follows an universal curve which depends only on the class of universality of the critical phenomenon.

\begin{figure}[h]
\includegraphics[width=8cm, clip=true]{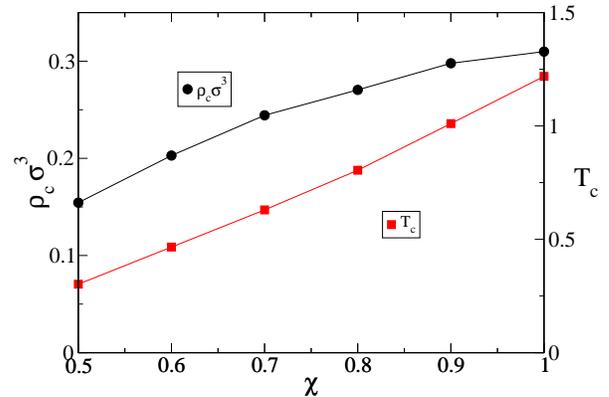}
\caption{Coverage  $\chi$ dependence of the critical density $\rho_c$ 
and temperature $T_c$.  }
\label{fig:tc-vs-coverage}
\end{figure}

The critical parameters, resulting from the grand-canonical simulations, are summarized in Table~\ref{table:cp}.  The last column indicates the size of the largest simulation box, which has been
progressively increased to compensate the decrease of the critical density and the 
associated shift towards smaller number of particles of the density fluctuations.
 Table \ref{table:cp} also reports the critical  chemical potential and the
value of the second virial potential $B_2^c$, normalized to the hard-sphere value $B_2^{HS}$
\begin{equation}
B_2^c/B_2^{HS} = 1- \chi^2 [(1+\frac{\Delta}{\sigma})^3-1] [\exp(\frac{u_0}{k_B T_c})-1].
\end{equation}
 also evaluated at the critical point.  Previous work\cite{FoffiKern} has shown that $B_2^c/B_2^{HS} $
 becomes smaller and smaller with decreasing valence. The actual value of 
  $B_2^c/B_2^{HS} $  can provide an  estimate of the effective valence of the system.  A comparison with Table 1 of Ref.~\cite{FoffiKern}  suggests that when $\chi$ reaches the value 0.6, the effective valence becomes  less than 4 and that when $\chi=0.5$ the effective valence is still larger than three. The critical parameters  $T_c$ and $\rho_c$  are also shown graphically in Fig.~\ref{fig:tc-vs-coverage}.  In agreement with the progressive reduction of the valence, $T_c$ and $\rho_c$  do decrease with decreasing $\chi$. 
Note that results presented in Figure \ref{fig:phase} have a counterpart in the case where the attractive part is spread over two patches 
distributed at the opposite poles of the sphere (see Fig.2 in Ref.\cite{kerntwo}).

\begin{table}[htdp]
\caption{Critical parameters as a function of the coverage. The last column indicates the size of the largest studied simulation box.}
\centering
\begin{tabular}{|c|c|c|c|c|c|c|}
\hline
$\chi$ & $\rho_c $ & $T_c$ & $\beta_c \mu_c$ & $ \mu_c $ &  $B_{2}^{c}/B_{2}^{HS}$ & L \\ 
\hline
\hline
1 & 0.31 & 1.22 & -2.955 & -3.601 & -2.020 & 7 \\
0.9 & 0.30 & 1.01 & -3.048 & -3.079 & -2.253 & 7 \\
0.8 & 0.27 & 0.800  & -3.270 & -2.613 & -2.790 & 11 \\
0.7 & 0.24 & 0.610 & -3.684 & -2.248 & -3.828 & 15 \\
0.6 & 0.20 & 0.446 & -4.482 & -1.200 & -6.187& 15 \\
0.5 & 0.15 & 0.302 & -6.371 & -1.924 & -14.68& 15 \\
\hline
\end{tabular}
\label{table:cp}
\end{table}%

When $\chi=0.5$ (the Janus case), a new interesting phenomenon appears. 
Fig.~\ref{fig:phase} shows indeed that, differently from the standard behavior (and for all the studied $\chi>0.5$ cases)  the density of the gas phase along the coexistence line increases progressively on cooling. This
peculiar behavior is  discussed in full details in the following sections.
For smaller values of $\chi$ we have not presently been  able to evaluate the phase diagram for two different reasons:
i) the temperature region where the critical point is expected (around $T=0.17$ for $\chi=0.4$, by a quadratic extrapolation of the data reported in Table~\ref{table:cp}) requires significant computational resources.
ii) preliminary tests have detected the formation of lamellar phases for $\chi=0.4$ already at $T>0.17$.
Despite the disappearance of the liquid phase as an equilibrium phase in the presence of  anisotropic potentials
is a potentially relevant issue\cite{newromano}, we can not at the present time address this point for the present model.
 A study of the stability of the liquid phase as compared to the (unknown) ordered phases  will be the argument of a future study, which will require the use of algorithms to identify the possible crystal structures\cite{genetic,marjolin} as well as free-energy evaluations to establish the relative stability compared to the liquid phase. A similar study for tetrahedral patchy particles
 has been recently reported\cite{newromano,romano2}. 
  Indeed, it could well be, as suggested in a recent study of particles with two patches\cite{kerntwo},  that the reduction of the  bonding angle could play a role analog to the reduction of the range in spherically interacting potential\cite{hagen_frenkel,Ile95a} and limited valence potentials\cite{newromano}, where the liquid phase disappears  when the range is smaller than $\approx 0.2 \sigma$. 

%% Achille: Forse dovremmo mettere una referenza per questo risultato, probabilmente W. G. Kranendonk and D. Frenkel, Mol. Phys, \textbf{64}, 403 (1988) 
 
%
%  a cos=-0.2 a T=0.25 e T=0.20 la fase densa mostra piani (crystal --- mgl files sul laptop). T=0.30 e' ok. 
%
%\begin{figure}[h]
%\includegraphics[width=8cm, clip=true]{zeta-betamu-mu-vs-covarage.pdf}
%\caption{ }
%\label{fig:muc}
%\end{figure}
%\begin{figure}[h]
%\includegraphics[width=8cm, clip=true]{tc-vs-rhoc.pdf}
%\caption{ }
%\label{fig:muc}
%\end{figure}
%\begin{figure}[h]
%\includegraphics[width=8cm, clip=true]{b2-tc-rhoc-vscosteta.pdf}
%\caption{ }
%\label{fig:muc2}
%\end{figure}

\section{Results: Janus}

%size11.png      size130.png     size22.png      size33.png      size63.png
%size117.png     size191.png     size29.png      size50.png      size79.png

\begin{figure*}[hbtp]
\begin{center}
\mbox{
	\leavevmode
	\subfigure [ s=11 ]
	{ \label{f:subfig-1}
	  \includegraphics[width=1.7cm, clip=true]{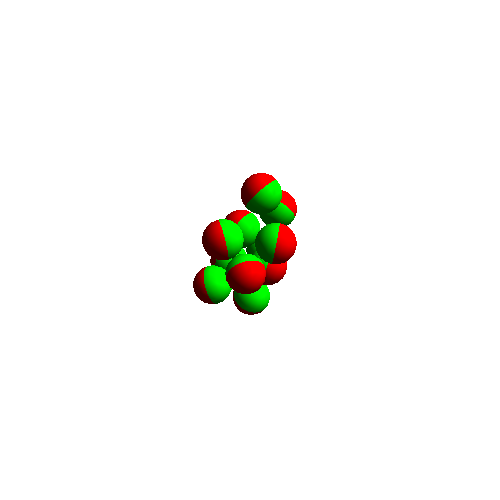} }

		\leavevmode
	\subfigure [  s=22  ]
	{ \label{f:subfig-2}
	  \includegraphics[width=1.7cm, clip=true]{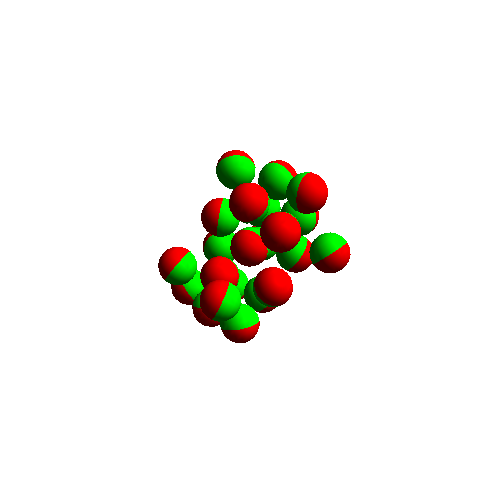} }
	  
	  \leavevmode
	\subfigure [ s=33  ]
	{ \label{f:subfig-3}
	  \includegraphics[width=1.7cm, clip=true]{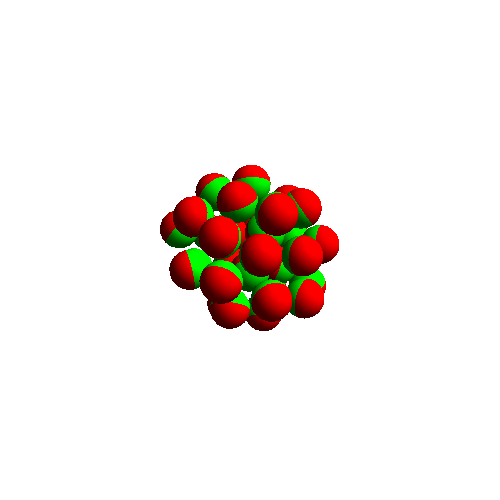} }

\leavevmode
	\subfigure [  s=50  ]
	{ \label{f:subfig-4}
	  \includegraphics[width=1.7cm, clip=true]{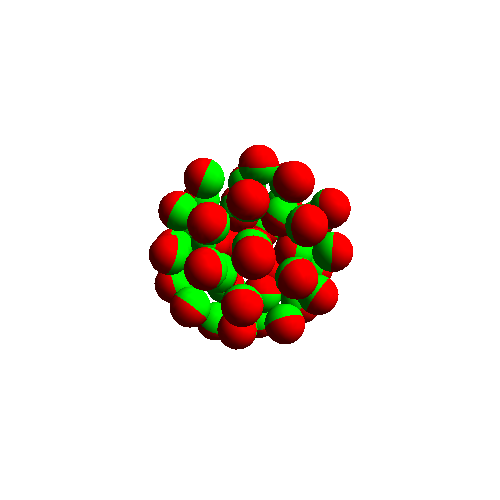} }

\leavevmode
	\subfigure [  s=63 ]
	{ \label{f:subfig-5}
	  \includegraphics[width=1.7cm, clip=true]{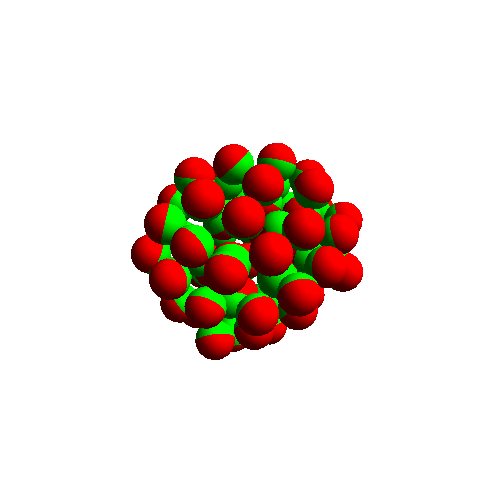} }
	  
	  \leavevmode
	\subfigure [  s=79]
	{ \label{f:subfig-5}
	  \includegraphics[width=1.7cm, clip=true]{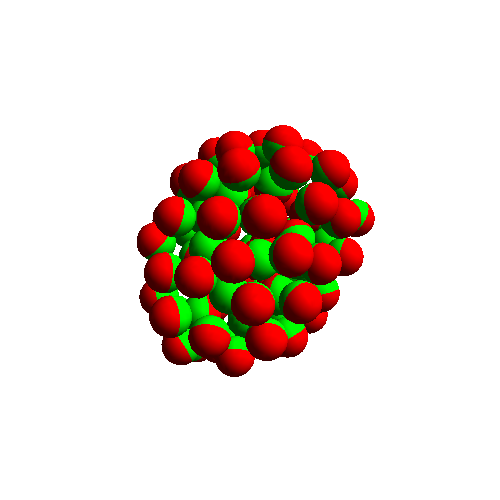} }
	  \leavevmode
	\subfigure [  s=117 ]
	{ \label{f:subfig-5}
	  \includegraphics[width=1.7cm, clip=true]{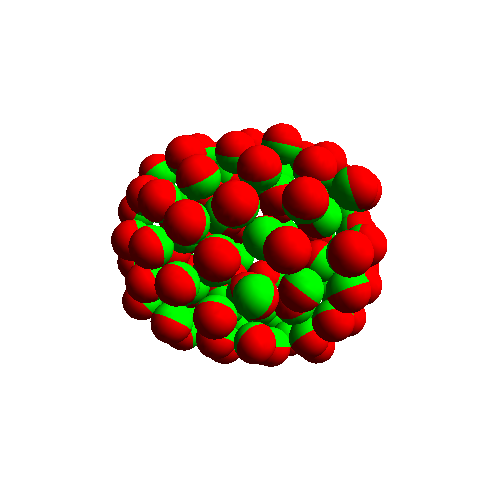} }
	  \leavevmode
	\subfigure [  s=130]
	{ \label{f:subfig-5}
	  \includegraphics[width=1.7cm, clip=true]{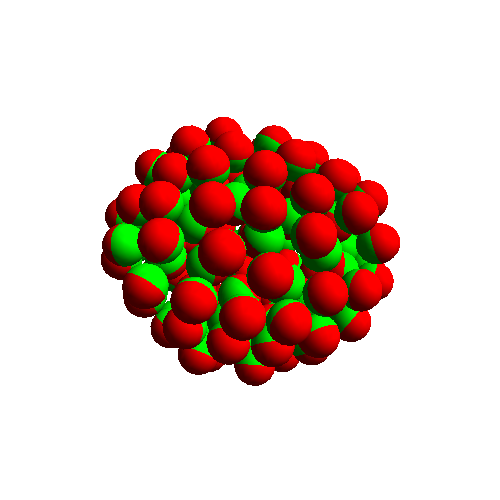} }
	  \subfigure [  s=191 ]
	{ \label{f:subfig-5}
	  \includegraphics[width=1.7cm, clip=true]{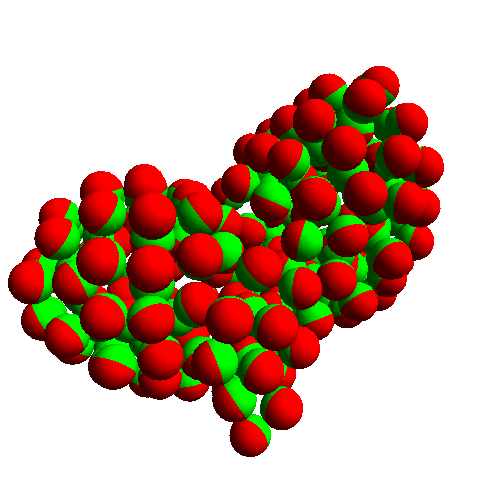} }
}	 

\end{center}

\caption{Typical cluster shapes of different size, extracted from simulations at $T=0.27$.}
\label{fig:snapshot}
\end{figure*}

\begin{figure}[h]
\includegraphics[width=9cm, clip=true]{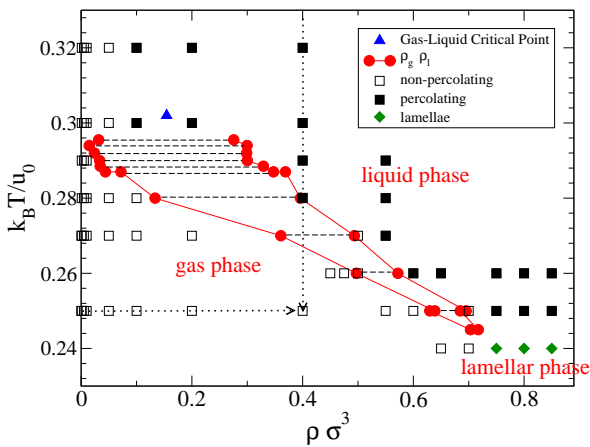}
\caption{Phase diagram of the Janus particles.  Filled (red) circles indicate the gas-liquid coexistence lines with the (blue) triangle denoting the critical point. The filled and open squares indicate the percolating and non-percolating state points, respectively, whereas (green) diamonds indicate the simulations that show a lamellar phase. Dashed lines connect coexisting state points whereas the two dotted lines refer to the two paths followed in
the calculation of the structure factor of Fig. \ref{fig:sq}.} 
\label{fig:points}
\end{figure}

The gas-liquid coexistence for the Janus ($\chi=0.5$) particles is enlarged in Fig.~\ref{fig:points}.
As discussed in a preliminary publication\cite{janusprl}, the phase diagram 
has a very odd behavior. The gas coexisting density, which typically is a decreasing function
of the temperature, here increases progressively on cooling, approaching the 
coexisting liquid density. In a simple liquid, coexistence between gas and liquid is
established on the basis of a compensation between the gas and liquid free energies.
The lower energy of the liquid phase is compensated by a larger entropy of the gas phase, which is acquired by significantly increasing the volume per particle.  
As discussed in Ref.~\cite{janusprl} (and detailed more in the following), 
 the uncommon behavior observed in the Janus case arises by a completely different 
 compensation mechanism between the liquid and the gas. The gas becomes the
 energetically stable phase (due to the formation of orientationally ordered aggregates,
 micelles and vesicles) and the liquid phase instead is stabilized by the larger orientational entropy  of the particles. This odd behavior give rises to a gas-liquid coexistence curve in the
 $P-T$ plane which is negatively sloped and to an expansion of the system on crossing from
 the gas to the liquid phase on cooling along isobars  (see Fig. 4 in Ref.\cite{janusprl}).
 
This anomalous thermodynamic behavior arises from the progressive establishment in the gas phase of clusters of particles which  --- due to the surface  pattern  properties of the particle ---  organize themselves in particularly stable structures.  Typical cluster shapes for different values of the cluster size $s$ are shown in Fig.~\ref{fig:snapshot}.  For small cluster sizes ($s \lesssim 20$) clusters are of micellar type, i.e. formed by aggregates in which the attractive part constitutes the
core of the aggregate. For larger size, the cluster organization changes in favor of a double layer structure, reminiscent of vesicles, in which the inner and outer surfaces are repulsive and the inner core is attractive.   Here, and in the rest of the manuscript, clusters are defined as set of particles connected by an uninterrupted path of bonds, where we define bonded any pair of particles whose pair potential energy is $-u_0$.

In this article we explore in details the properties of the clusters which develop in the gas phase, 
their structure,  energy and abundance. We then investigate the gas and liquid phases with the aim of characterizing the collective structure of the system  (both in real and wave-vector space) as well as its connectivity properties
(percolation).

%% Achille: Ho cambiato la figura in modo che fosse piu' semplice da leggere e fosse coerente con il PRL.

\subsection{Critical Micelle Concentration}

\begin{figure}[h]
           \includegraphics[width=8.cm, clip=true]{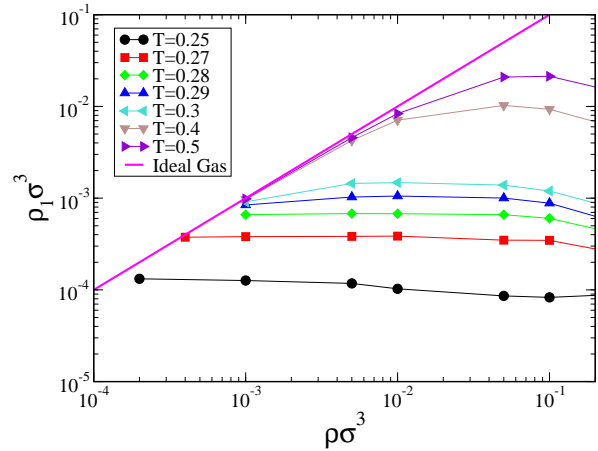}
\caption{Relation between the monomer number density $\rho_1$ and the system number density $\rho$ at different $T$. The flat region for  $T \lesssim 0.28$ indicates that the addition of monomers preferentially results in the formation of aggregates. The intercept between the flat curve and the ideal gas behavior provides an accurate estimate of the critical micelle concentration.}
\label{fig:rho1}
\end{figure}

The onset of  micelles can be demonstrated and quantified by the study of the relation between the density  $\rho_1$  of particles in monomeric state (i.e.  un-bonded  particles) and the system density $\rho$. Fig.~\ref{fig:rho1} shows that  for $T \lesssim 0.28 $, a sharp kink separates the ideal gas  behavior   (where all the particles are in a very dilute monomeric state and $\rho_1=\rho$) from a rather insensitive density dependence of the number of monomers in solutions, a clear indication that, at low $T$, the addition of particles to a constant volume system promotes the formation of additional aggregates.  This behavior is indeed typical of micelle forming systems\cite{bookcolloid}, and the location of the kink provides an estimate of the critical micelle concentration (c.m.c.), which in the present case varies from $\rho=10^{-3}$ down to $\rho=10^{-4}$ when $T$ changes from $T=0.28$ to $T=0.25$.

One may wander why this system, on cooling, does not show a typical gas-liquid coexistence. In simple fluids, at low $T$, the gas phase does not show significant clustering. Indeed, on increasing the density, the system transforms into a liquid phase, by establishing an infinite size percolating clusters, in the attempt to  minimize the potential energy, being the entropic loss in the free energy (associated to the restricted sampling of the  system available volume) made less relevant  by the small $T$.  In the present case, the Janus potential allows for the establishment of significantly bonded aggregates, i.e. with a small potential energy, without the need of forming an infinite size cluster.  By exposing the hard-core part to the exteriors, these clusters do not feel any driving force toward further clustering.

\subsection{Cluster Size Distributions,  percolation}
\label{sec:csd}
\begin{figure}[h]
\includegraphics[width=7.5cm, clip=true]{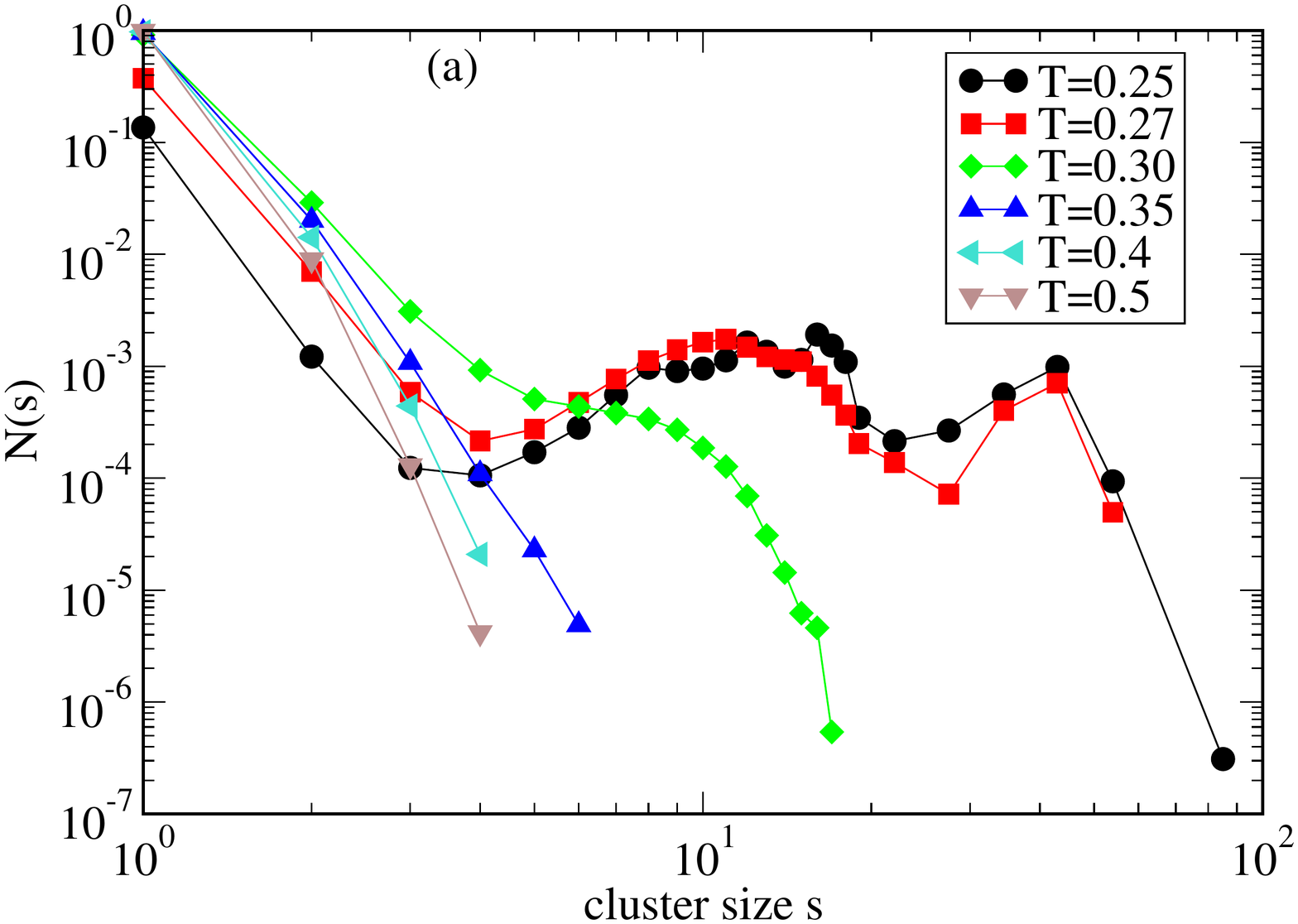}
\includegraphics[width=7.5cm, clip=true]{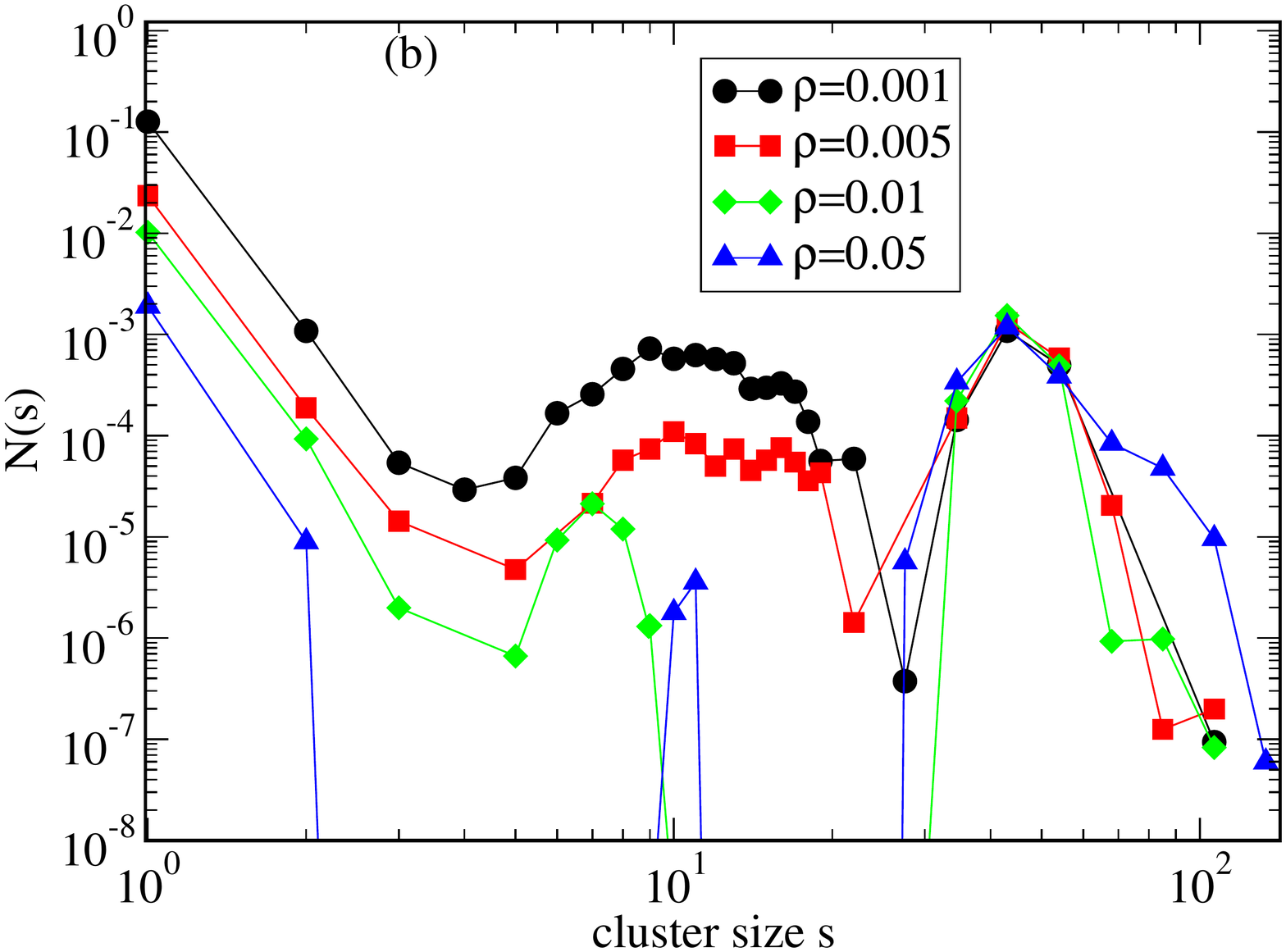}
\caption{Cluster size distributions at $\rho=0.001$ for several $T$ values  (a) and at $T=0.25$ for several densities (b). The distributions have been normalized so that $\sum_s s N(s)=1$.}
\label{fig:csd}
\end{figure}

To quantify the effect of clustering and its $\rho$ and $T$ dependence we  show in 
Fig.~\ref{fig:csd} the cluster size distribution, $N(s)$.

Below the critical temperature, the gas phase becomes populated by rather stable
clusters. Particularly significant are clusters of size between 10-15 (micelles)
and clusters between size 40 and 50 (vesicles). Other sizes are also found,
but their statistical relevance decreases on decreasing $T$. 
This feature appears clearly in the cluster size distribution, $N(s)$, shown in 
Fig.~\ref{fig:csd}. On decreasing $T$, the monotonic decaying $N(s)$ 
develops a shoulder around $T_c$, which evolves into a clear 
two-peaked function on further cooling, signaling the appearance of
micelles and vesicles (Fig.~\ref{fig:csd}-(a)). 
Part (b) of Fig.~\ref{fig:csd} shows the evolution of $N(s)$ on increasing
$\rho$.  The micelle peak progressively empties in favor of the vesicles peak, which
at low $T$ and large $\rho$ are the most stable structures.
Similar trends are seen at other $T$ or $\rho$.

\begin{figure}[h]
\includegraphics[width=9cm, clip=true]{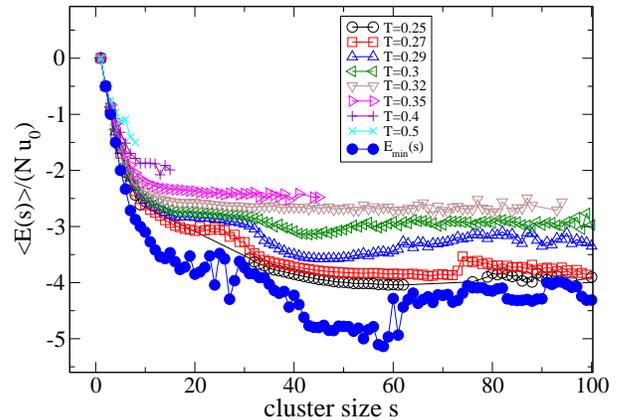}
\caption{Average potential energy (per particle) of clusters of different size $s$ at several $T$ at $\rho=0.01$. For each $s$, it is also shown the lowest energy ever observed $E_{min}(s)$, independently from $T$ or $\rho$.}
\label{fig:esize}
\end{figure}

An indication of the relative stability of the clusters as a function of their size
is provided by the cluster potential energy.  Fig.~\ref{fig:esize} shows the
average potential energy $\langle E(s) \rangle$, averaged over all clusters of the same size, vs. size $s$ for different $T$ at fixed $\rho$. One notices that for each size there are 
several distinct arrangements with different energies and entropies. On cooling,
lower energy clusters become preferentially selected and 
$\langle E(s) \rangle$ decreases. One also notices a plateau  between size 10 and 30
followed by an additional plateau, at low $T$, starting from $s \approx 40$.
The initial location of the plateaus coincides with the size of the  mostly represented clusters. This can be understood 
by considering that the free energy of the system has both  energetic and entropic contributions.
When the potential energy per particle does not change with the size of the cluster, then
it becomes convenient to the system to favor the formation of the smallest possible cluster with the same energy,
since it will be possible in this way to maximize the total number of clusters and hence the
translational component of the entropy. %, as discussed in more details below.

Fig.~\ref{fig:esize}  also shows the lowest energy configuration $E_{min}(s)$ found
for each cluster size, independently from temperature and density of the
simulation. This curve provides an estimate for the cluster ground state and
clearly show that the lowest-energy structures are the vesicles, i.e. clusters composed
between 40 and 60 particles.

We have examined the connectivity properties of the different state points by evaluating the 
presence of clusters spanning the entire simulation box.  When more than 50 per cent of the
configurations are characterized by spanning clusters, the state point is classified as
percolating.   Fig.~\ref{fig:points} shows the location of the percolating state points in
the phase diagram. The cluster gas phase is never percolating, while the
liquid states always are, so that percolation properties can also be used to 
distinguish between the two phases. We also confirm that the critical point is located inside the
percolation region, confirming once more that a pre-requisite for  critical phenomena is the
existence of a percolating network of interactions\cite{coniglio}.

\subsection{Structure Factor}

\begin{figure}[h]
\includegraphics[width=7.8cm, clip=true]{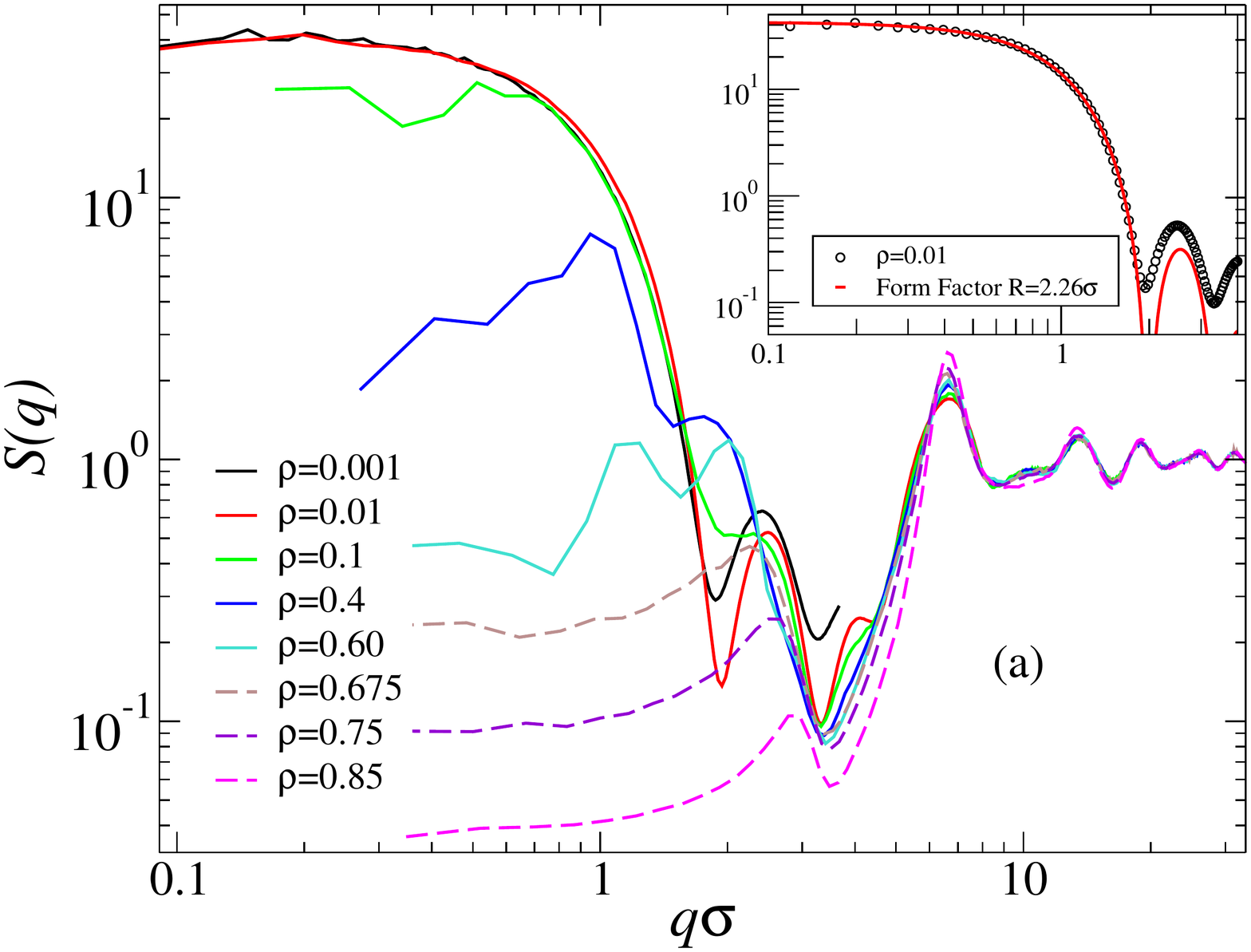}
\includegraphics[width=7.5cm, clip=true]{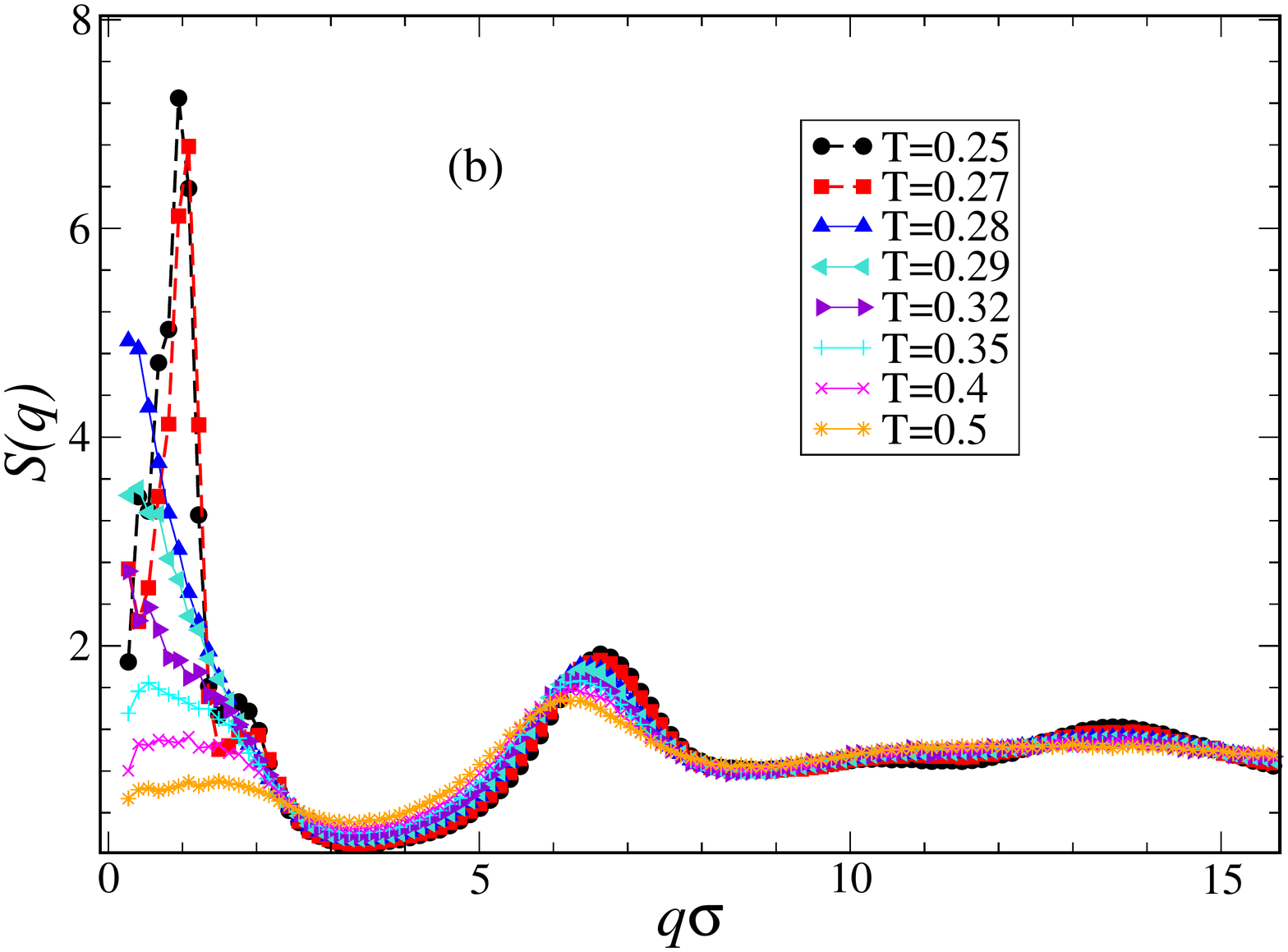}
\caption{Structure factor at (a) T=0.25 for several densities and at (b) $\rho=0.4$ for several  $T$. The inset in (a) shows the fit with the form factor of a sphere (Eq.~\ref{eq:formf}). The best-fit
value for the radius is $R=2.26 \sigma$.  The above isothermal and isochore paths are indicated with dotted lines in Fig.\ref{fig:points} } 
\label{fig:sq}
\end{figure}

It is particularly relevant to look at the structure factor in this system both in
the gas phase at low $T$, when micelles and vesicles become prominent, 
as well as in the liquid phase. 
Fig.~\ref{fig:sq}-(a) shows the evolution of the structure factor on increasing density
at  $T=0.25$, across the gas-liquid transition. At this low $T$, as discussed in Sec.~\ref{sec:csd},
vesicles are the dominant clusters. The structure factor indeed evolves from the
one characteristic of an ideal gas of spherical vesicles (and indeed the inset shows that $S(q)$ is 
well represented by the form factor $P(q)$ of a sphere of radius $R$)
\begin{equation}
P(q)= \left[  \frac{3(\sin(qR)-qR\cos(qR))}{(qR)^3} \right]^2,
\label{eq:formf}
\end{equation} 
to the one of interacting spheres, in the micelle-rich gas phase. Indeed, on increasing $\rho$, oscillations at $q\sigma \approx 1.2$ arise, which correspond to distances comparable to the vesicle size.  
Beyond density $\rho=0.6$ the gas condenses into the liquid phase, where only a very weak pre-peak around $q\sigma \approx 3$ is found.

%% Achille: qui non capisco bene la cosa. A $q\sigma \approx 1.2$ corrisponde ($q=2\pi/d$) $d/\sigma \approx 5.2$ e queste si dovrebbero
%% riferire alle dimensioni delle vesciole dato che non sono interagenti a questa densita'. Per il caso
%%  $q\sigma \approx 3$ si ottiene $d/\sigma \approx 2.1$ ma questa dovrebbe essere la distanza tra i primi vicini nel liquido?

Fig.~\ref{fig:sq}-(b) shows the evolution of the structure factor on cooling  along the $\rho=0.4$ isochore.
Compared to simple liquids, one observes a non negligible scattering at small $q$, which progressively
increases on approaching the phase separation. These are the standard critical fluctuations which are
expected to diverge on approaching a spinodal line.  While in simple liquids, 
below the spinodal temperature the system phase separate in a  gas coexisting with a liquid phase,
here, the peculiar shape of the gas-liquid coexistence line (Fig.~\ref{fig:phase}) opens up new
stable states, composed by interacting vesicles and $S(q)$ becomes peaked at the
vesicle-vesicle distance.

\subsection{Angular Correlations}

\begin{figure}[h]
\includegraphics[width=7cm, clip=true]{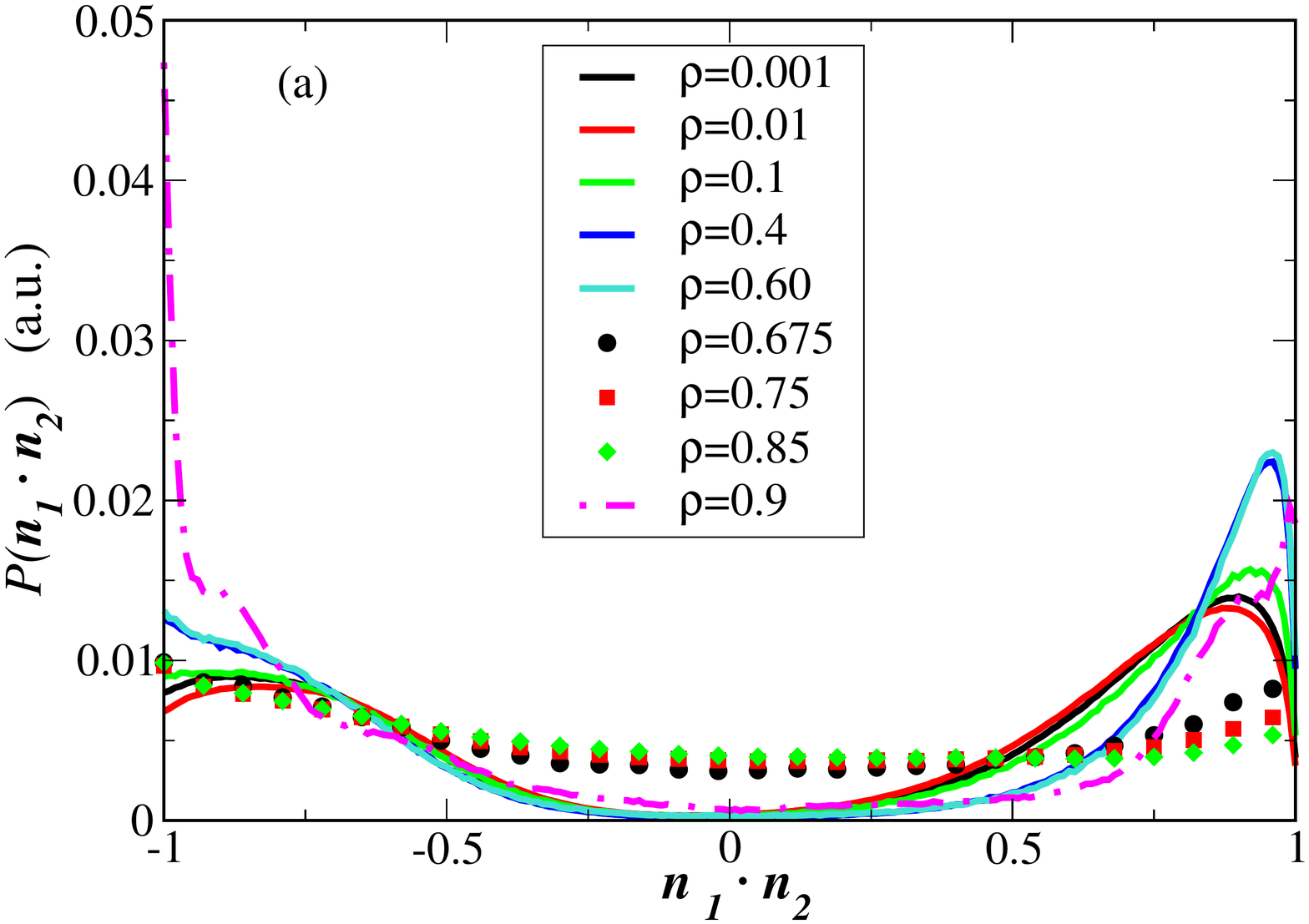}
\includegraphics[width=8cm, clip=true]{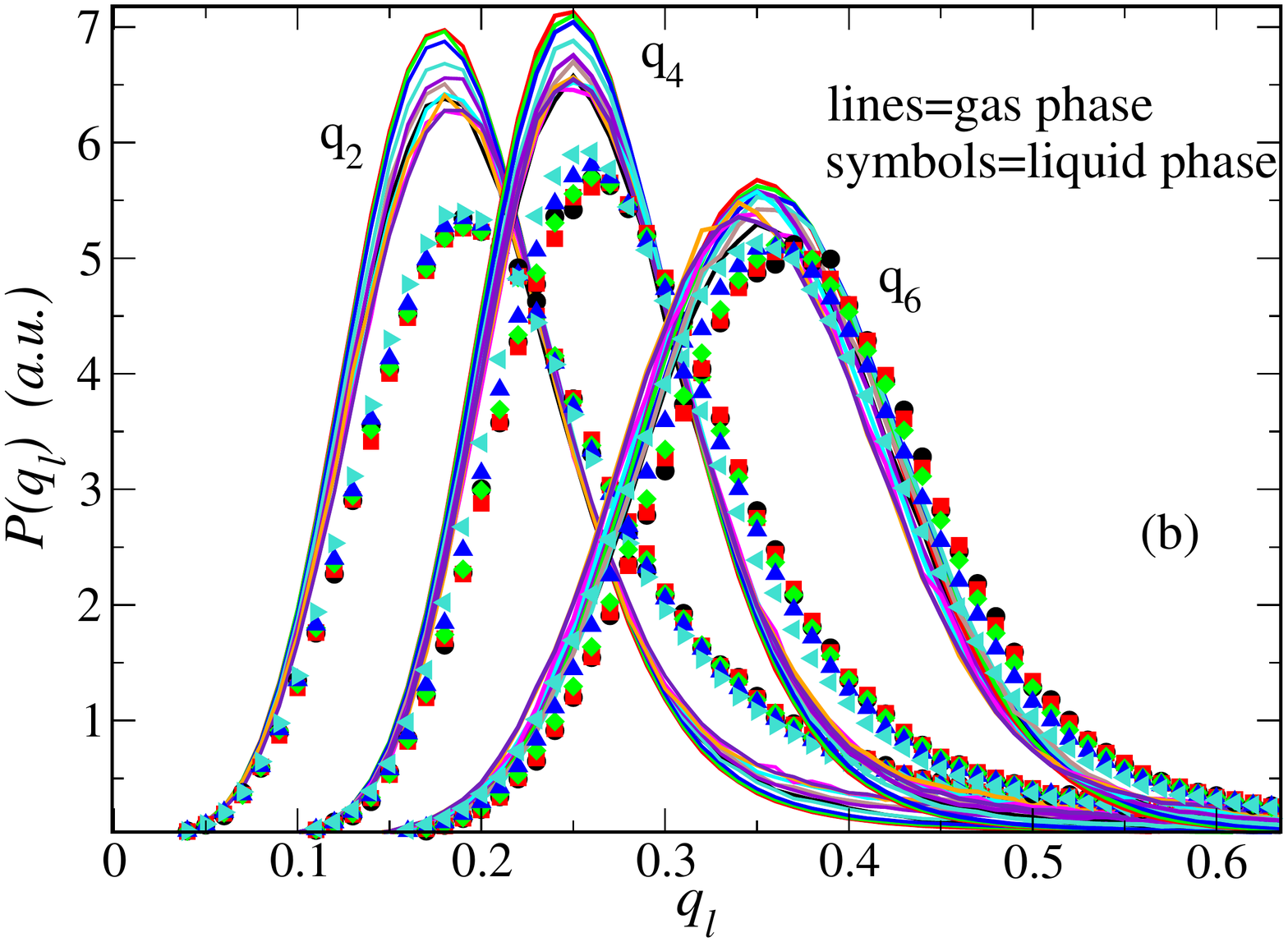}
\caption{(a) Distribution of the scalar product between the unit vectors  of all bonded pairs of particles. The gas-like
state points are characterized by peaked distributions, revealing the presence of orientational order. The vesicle configurations significantly contribute to the region ${\bf n_1 \cdot n_2} \approx 1$, while the micelle configurations contribute to  ${\bf n_1 \cdot n_2} \approx -1$.
%The relative orientation corresponding to the scalar products $-1$, $0$ and $1$ are also depicted. 
(b) Distribution of the rotational invariant $q_l$ in the gas and in the liquid phase. }
\label{fig:costeta}
\end{figure}

%% Achille: anche qui ho cambiato la notazione dei versori in modo da essere omogenei. Ho anche cambiato la Figura 9 ma non sono ancora 
%%riuscito a fare i versori. Nella figura ho inserito le sferette perche' altrimenti e' faticoso da leggere.
To provide evidence of the different ordering of the particles in the gas and in the liquid phase,
we have  calculated the distribution of relative orientations between all pairs of bonded particles. More precisely, we have evaluated the distribution of the scalar product ${\bf \hat{n}}_1 \cdot {\bf \hat{n}}_2$ where ${\bf \hat{n}}_1$ and ${\bf \hat{n}}_2$  are the two  {unit vectors} indicating the location of the patch center in each particle frame, for all bonded pairs. 
The distribution $P({\bf \hat{n}_1} \cdot {\bf \hat{n}_2})$ is expected to show well defined peaks
for an ordered state and to be flat in a completely disordered  state.
Fig.~\ref{fig:costeta} shows such distribution for several $\rho$ values at low $T$.  The orientational
differences between the ordered gas phase and the disordered liquid phase appear very clearly. 
On increasing $\rho$, $P({\bf \hat{n}_1} \cdot {\bf \hat{n}_2})$ evolves from a highly structured function with peaks
close to $| {\bf n}_1 \cdot {\bf \hat{n}}_2 | >0.5$   in the  vesicle-rich gas phase, to a 
more uniform distribution for the liquid phase state points, going back to an evenly more structured 
function when the lamellar phase is entered. 
The peak at ${\bf \hat{n}}_1 \cdot {\bf \hat{n}}_2$ close to $-1$ is created by the pairs of particle which face each other (as in a double-layer), while the peak at  ${\bf \hat{n}}_1 \cdot {\bf \hat{n}}_2$ close to $0.9$ arises from particles which are forming the shells of the vesicle.

We note that, to quantify the orientational ordering in the system, it is necessary to identify an order parameter which
does not depend on single-particle properties, since both gas and liquid phases are isotropic. We have also
investigated the rotationally invariant local order\cite{Stein83} indicators, which have been often 
exploited to quantify  order   in crystalline solids, liquids
and colloidal gels\cite{Cam05aPRL,Sciobartlett}.

\begin{equation}
q_{l}(i) \equiv  \left[\frac{4 \pi}{2l+1} \sum_{m=-l}^{l} |
\bar q_{lm}(i)|^2 \right]^{1/2}
\end{equation}
where $\bar q_{lm}(i)$ is defined
as,
\begin{equation}
\bar q_{lm}(i) \equiv \frac{1}{N_{b_i}}
\sum_{j=1}^{N_{b_i}}  Y_{lm}(\mathbf{\hat{r}}_{i j})
\end{equation}
Here $N_{b_i}$ is the set of bonded neighbors of a particle $i$.  The
unit vector $\mathbf{\hat{r}}_{ij}$ specifies the orientation of the bond
between particles $i$ and $j$. In a given coordinate frame, the
orientation of the unit vector $\mathbf{\hat{r}}_{ij}$ uniquely determines the
polar and azimuthal angles $\theta_{i j}$ and $\phi_{ij}$.  The
$Y_{lm}(\theta_{i j},\phi_{ij}) \equiv Y_{lm}(\mathbf{\hat{r}}_{i j})$ are the
corresponding spherical harmonics. 
We have  calculated the distribution of $q_{l}$ for the present model. As shown in Fig.~\ref{fig:costeta}-(b), 
the  resulting distributions  show systematic differences between gas and liquid phase, but it is hard to provide a clear connection with the degree of order in the system.

\subsection{Pathways for vesicles formation}

\begin{figure*}[hbtp]
\begin{center}
\mbox{
	\leavevmode
	\subfigure [ 25.000 ]
	{ \label{f:subfig-1}
	  \includegraphics[width=2.3cm, clip=true]{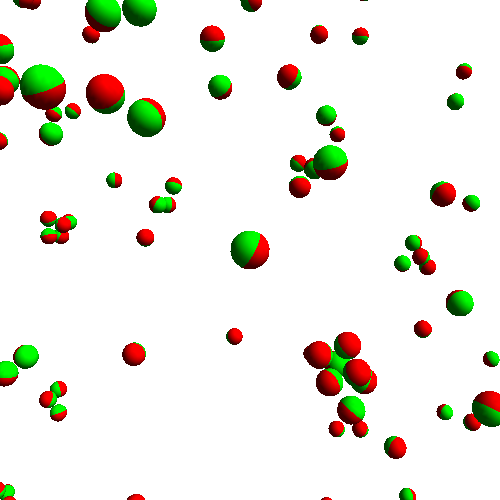} }

		\leavevmode
	\subfigure [  125.000  ]
	{ \label{f:subfig-2}
	  \includegraphics[width=2.3cm, clip=true]{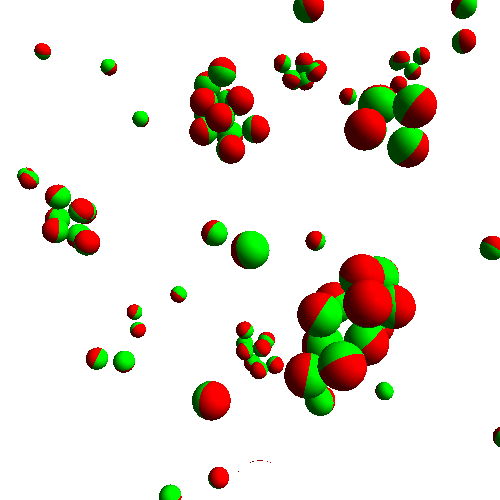} }

	  \leavevmode
	\subfigure [  500.000 ]
	{ \label{f:subfig-5}
	  \includegraphics[width=2.3cm, clip=true]{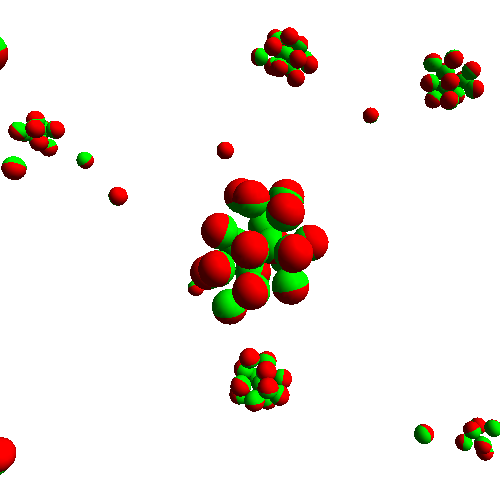} }

	  \leavevmode
	\subfigure [ 750.000  ]
	{ \label{f:subfig-3}
	  \includegraphics[width=2.3cm, clip=true]{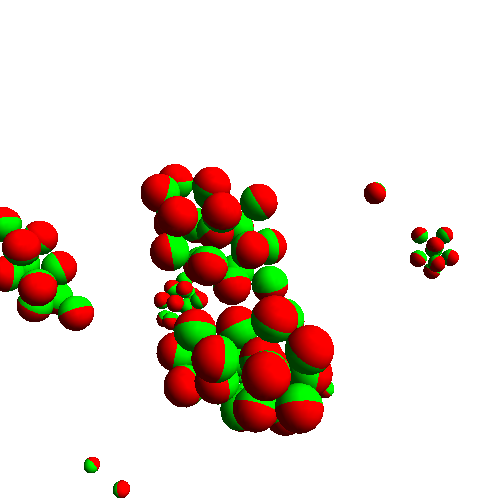} }

\leavevmode
	\subfigure [  825.000  ]
	{ \label{f:subfig-4}
	  \includegraphics[width=2.3cm, clip=true]{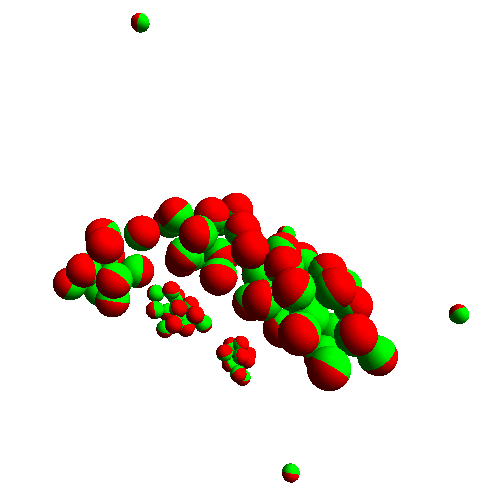} }

	  \leavevmode
	\subfigure [ 950.000]
	{ \label{f:subfig-5}
	  \includegraphics[width=2.3cm, clip=true]{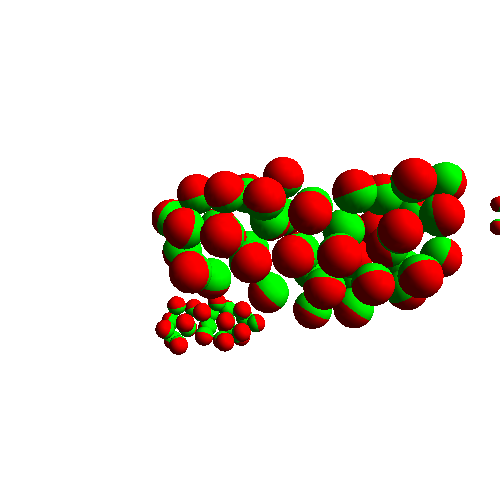} }
	  \leavevmode
	\subfigure [  1.000.000 ]
	{ \label{f:subfig-5}
	  \includegraphics[width=2.3cm, clip=true]{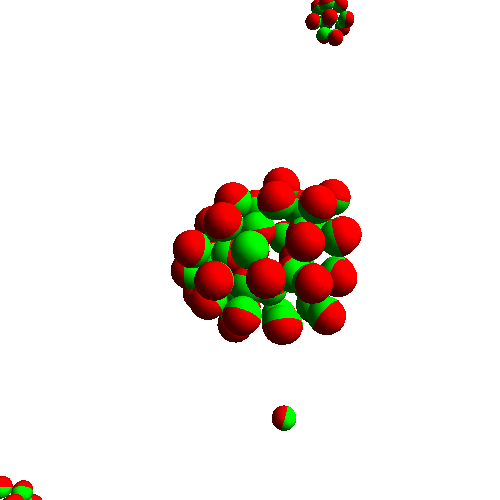} }
	  \leavevmode
	}	 

\end{center}

\caption{Snapshot from a simulation at $\rho=0.005$ and T=0.25 for several MC steps, indicated in the labels. 
The initial configuration, composed by isolated monomers, quickly evolves to form small micelles (a-c).
The final vesicle (g) arises from a collision between three distinct micelles (d-e) which form an elongated transient tubular cluster (f). Each picture has a side length of $9 \sigma$.}
\label{fig:formazione}
\end{figure*}

Despite Monte Carlo simulations do not allow for a precise definition of time, it is interesting to analyze how 
particles rearrange themselves into large aggregates. More specifically, the analysis of the sequence of
MC configurations may help understanding how vesicles are generated.  We have examined graphically
several processes of formation of a vesicle starting from a monodisperse solution of particles and one
of these processes is documented in Fig.~\ref{fig:formazione}. In all cases, the system starts by forming small clusters that grow by incorporating isolated monomers or
colliding to similar clusters, to form micelles (Fig.~\ref{fig:formazione}-(a-c)). The transition to vesicle requires the interaction between
three micelles (d-e), with the formation of a long cylindrical micelle (f) which then self-restructures itself into the more
energetic vesicles configuration (g).

\section{Conclusions}

This article reports a numerical study of a simple potential for Janus-like particles, i.e. colloidal spherical particles whose surface is divided evenly into two areas  of different chemical composition.  This study is motivated by the ongoing 
effort in the direction of synthesizing optimal Janus colloids\cite{Janus,janusgold, janusweitz,Stellacci,granich}. The
possibility of creating particles whose surface has different behavior on the two hemispheres significantly
enlarges the richness of the resulting collective behaviors. The phase diagram, the ordered and disordered 
stable structures, the self-assembly properties of optimal clusters are indeed expected to finely depend on the
chemistry and physical properties of the particle surface.  For this reason, the assembly behavior of Janus particles is receiving a considerable attention even from a theoretical and numerical point of view\cite{cacciuto, phasejanus,cacciuto2,vanakaras,whitelam}.   With proper choices of the  chemico-physical  surface properties, Janus particles can provide the most elementary and geometrically simple example of a surfactant particle\cite{erhardt}, in which solvophilic and solvophobic  areas reside on different part of the surface of the same particle.   

Modeling of the phase behavior of these particles can be performed  at different levels of realism. In this article
we have chosen to implement the highest coarse-graining procedure, by considering the two sides  of the particle as
repulsive and attractive,  and modeling each of them with the simplest corresponding potential,  a hard-core plus
a square-well.  Despite this strong simplification, the phase diagram of the system 
 displays a very rich behavior, with a colloidal-poor (gas) colloidal-rich (liquid) de-mixing region, which is progressively suppressed by the insurgence of micelles, thus providing a model where micelle formation and phase-separation are   simultaneously observed.  The study of this model shows that the suppression of the 
phase separation is driven by the possibility of building low energy clusters   which are 
shielded by the presence of an external hard-sphere surface, diminishing the driving force for forming 
large-size aggregates. As a result,  at low $T$, the system organizes  into a dispersion of orientational ordered micelles and vesicles, essentially interacting via excluded volume only.  

Another  advantage of studying the one-patch Kern-Frenkel potential is that the parameters
of the potential can be tuned  continuously from the square-well to the Janus potential, simply
by changing the bonding angle, that is the coverage $\chi$.  This has made possible to follow the evolution of the 
gas-liquid coexistence curve with $\chi$,  thus including the Janus case
which displays the inherintly interesting simultaneous presence of a critical point and of a micelle formation.  
It would be interesting in the future to explore more in details the transition from the standard gas-liquid coexistence 
to the  more exotic Janus case exploring a more refined grid of $\chi  $ values.  It is of course possible that
the stabilization  of micelle and vesicles can take place already at $\chi  =0.6$ but in a low $T$ region
where numerical simulations can not be performed at the present time. 

Several questions are still open including  the role of the interaction range. 
Experimentally realized Janus particles  are characterized by interaction ranges 
of the order of $0.1\sigma$ or smaller. The range width is an important variable, since
it controls both the internal flexibility of the aggregates (and hence the cluster entropy) and the  geometries of the lowest-energy aggregates (cluster energy).  In this respect, it can not be a priori foreseen if
micelle and vesicles will remain the  most stable clusters on further decreasing  the range.  We are currently repeating our study for smaller ranges, but this  requires a major computational effort, since
 clustering shifts to lower  temperature, making studies of the
 phase diagram (in particular Gibbs ensemble calculations) extremely hard with present time numerical resources.  
 The sensitivity of the cluster shape to the cluster range would be an important observation and could teach us
 a lot about how to control the shape of the assembly experimentally, and, perhaps, help us understanding
 geometric arrangements which are found in protein aggregates.   In this respect it is  worth noticing that  micelles have been detected  in the most recent experimental work of Granick\cite{granich} together with  Bernal-spiral clusters, which have not been observed with the present model, perhaps due to differences in the interaction ranges.

\section{Acknowledgements}

FS acknowledges support from  ERC-226207-PATCHYCOLLOIDS, ITN-234810-COMPLOIDS and NoE SoftComp NMP3-CT-2004-502235.  AG acknowledges support from PRIN-COFIN 2007B57EAB(2008/2009).
We thank C. De Michele  for providing us with the code for generating  Fig.~\ref{fig:snapshot} and  Fig.~10, J. Horbach,  J. Largo, J. Russo and M. Noro for discussions.
% and ITN-234810-COMPLOIDS. 

\bibliographystyle{apsrev}
\bibliography{pccp-supra}

\end{document}